\documentclass[cits]{PoS}
\usepackage{amsmath}
\usepackage{subcaption}

\title{Radial and orbital excitation energies of charmonium}
\ShortTitle{Radial and orbital excitation energies of charmonium}

\author{HPQCD Collaboration}

\author{\speaker{B.\ A.\ Galloway}, P.\ Knecht, J.\ Koponen, C.\ T.\ H.\ Davies\\
        SUPA, School of Physics and Astronomy, University of Glasgow, Glasgow, G12 8QQ, UK\\
        E-mail: \email{b.galloway.1@research.gla.ac.uk}}

\author{G.\ P.\ Lepage\\
Laboratory for Elementary-Particle Physics, Cornell University, Ithaca, NY 14853, USA}

\abstract{The charmonium system has several excited states below the energy threshold for decay into $D$ and $\bar{D}$ mesons, which can in principle be studied accurately in lattice QCD. Studies that include many states in the spectrum have typically only been done at one value of the lattice spacing and with relatively heavy light quarks in the sea. Here we give preliminary results for radial and orbital excitation energies for charmonium from a calculation on 2+1+1 MILC configurations at multiple lattice spacings and including physical values for $u/d$ quark masses. We use the HISQ formulation for $c$ to obtain small discretisation errors and smeared operators to improve excited state overlap.}

\FullConference{The 32nd International Symposium on Lattice Field Theory\\
                 23-28 June, 2014\\
                 Columbia University, New York, NY}

\begin{document}

\section{The Charmonium System}
The system of mesons formed of a charm quark and its antiquark is known as charmonium. The energy states of this system are well-determined experimentally \cite{charmonium-pdg} and there are a number of states below the `open charm' threshold for decay into $D\bar{D}$. These states can be determined accurately using straightforward lattice QCD calculations with single meson operators. Previous lattice studies of many states in the spectrum (see \cite{dublin} for a particularly impressive example) have typically only been done at one lattice spacing, and/or with unphysical masses for light quarks in the sea. With only one lattice spacing, it is not possible to be sure of the value in the continuum limit, and so studies at multiple lattice spacings are required.

One such study is that by the Fermilab Lattice and MILC Collaborations \cite{milc2s1s}. However, this produced preliminary results for the spin-averaged 2S--1S splitting which showed a surprisingly large disagreement with the experimental value. This provided motivation for us to build on our earlier accurate calculation of ground state charmonium masses \cite{gordon} and determine whether we could also observe such a discrepancy. Here we present preliminary results from our study with physical light sea quarks at multiple lattice spacings.

\section{Lattice Calculation}
Two-point meson correlators were calculated using the MILC code, with the HISQ action \cite{hisq} used for the valence quarks to give small discretisation errors. Gaussian covariant smearings of the following form were applied to the source and sink operators to improve overlap with excited states:
	\begin{equation*}
		\left[ 1 + \frac{r_0^2 \cdot D^2}{4 \cdot n} \right]^n \xrightarrow{n\to\infty} \exp \left( \frac{r_0^2 \cdot D^2}{4} \right)
	\end{equation*}
where $n$ is the number of iterations, and the $r_0$ parameter determines the width of the Gaussian. $D$ is the stride-2 covariant difference operator, necessary because we are working with staggered quarks and must preserve the `taste' of the meson we are studying. Pairing up propagators with the smearings in Table \ref{smearings} applied to different combinations of the sources and sinks results in a matrix of correlators being obtained.

	\begin{table}
	\begin{center}
	\begin{tabular}{c|cccc}
				& Smearing 1	& $n_1$	& Smearing 2	& $n_2$ \\
		\hline
		Coarse	& 1.5			& 10	& 3.0			& 20 \\
		Fine	& 2.5			& 20	& 3.5			& 30 \\
	\end{tabular}
	\caption{Parameters for the Gaussian covariant smearings applied to the source and sink operators at different lattice spacings.}
	\label{smearings}
	\end{center}
	\end{table}

\subsection{Gauge Configurations}
We use configurations from the MILC collaboration with 2+1+1 flavours of HISQ quarks in the sea \cite{milc211}. Several ensembles exist with a wide range of lattice spacings --- we label these `very coarse', `coarse', `fine', and `superfine'. Details of the particular configurations used are shown in Table \ref{configs}. The ensembles utilised include physical light sea quarks, which means we can obtain results which require little or no chiral extrapolation.

	\begin{table}
	\begin{center}
	\begin{tabular}{lllllll}
		Label       & $a$ / fm & $m_{\ell}/m_s$ & Lattice size & $a m_c$ & $N_{\mathrm{cfg}} \times N_t$ & $w_0/a$ \\ 
		~           & (approx.)        & ~         & ($L^3 \times T$)  & ~       & ~       & ~ \\
	\hline
	\hline
		very coarse & 0.15             & 1/5       & $16^3 \times 48$  & $0.888$   &  $1020 \times 8$ &  $1.1119(10)$    \\
		~           & ~                & 1/10      & $24^3 \times 48$  & $0.873$   &  $1000 \times 8$ &  $1.1272(7)$    \\
		~           & ~                & phys      & $32^3 \times 48$  & $0.863$   &  $1000 \times 8$ &  $1.1367(5)$    \\
	\hline
		coarse      & 0.12             & 1/5       & $24^3 \times 64$  & $0.664$   &  $1053 \times 8$ &  $1.3826(11)$    \\
		~           & ~                & 1/10      & $32^3 \times 64$  & $0.650$   &  $1000 \times 8$ &  $1.4029(9)$    \\
		~           & ~                & phys      & $48^3 \times 64$  & $0.643$   &  $1000 \times 8$ &  $1.4149(6)$    \\
	\hline
		fine        & 0.09             & 1/5       & $32^3 \times 96$  & $0.450$   &  $300 \times 8$  &  $1.9006(20)$   \\
		~           & ~                & 1/10      & $48^3 \times 96$  & $0.439$   &  $300 \times 8$  &  $1.9330(20)$   \\
		~           & ~                & phys      & $64^3 \times 96$  & $0.433$   &  $565 \times 8$  &  $1.9518(7)$   \\
	\hline
		superfine   & 0.06             & 1/5       & $48^3 \times 144$ & $0.274$   &  $333 \times 4$  &  $2.8960(60)$   \\
	\hline
	\end{tabular}
	\caption{Details of the MILC configurations \cite{milc211} used in our calculations, along with the bare charm quark mass used, and the $w_0/a$ value determined, on each.}
	\label{configs}
	\end{center}
	\end{table}

Large numbers of configurations with several time sources have been used for each ensemble to provide high statistics. The bare charm masses we use are well-tuned by fixing to the $\eta_c$ mass and are also listed in Table \ref{configs}.

\subsection{Correlator Fits}
We perform Bayesian multi-exponential fits to the matrices of correlators that we obtain, using the \texttt{corrfitter} library designed for this purpose \cite{corrfitter}. We fit with up to $n=9$ exponentials, using a fit function of the form:
	\begin{equation*}
		\sum_{i=0}^{n-1} A_{i,\mathrm{sc}}A_{i,\mathrm{sk}} (e^{-E_it} + e^{-E_i(L_t-t)}) - (-1)^{t/a} \cdot B_{i,\mathrm{sc}}B_{i,\mathrm{sk}} (e^{-E_{o,i}t} + e^{-E_{o,i}(L_t-t)})
	\end{equation*}
where $E$ represents the energy of the fitted state, $L_t$ is the time extent of the lattice, and $t$ is the time between the source and sink of the correlation function (to which we fit). $A$ and $B$ are amplitudes, one for each different smearing (including no smearing) at the source (sc) and the sink (sk).
	
The use of the HISQ action for our valence quarks means we have an oscillating part in our vector correlators (accounted for by the second term in the fit function) which allows for access to axial vector states such as the $h_c$. The fits are all very good, with $\chi^2<1$ in each.

\subsection{Fixing the Lattice Scale}
To fix the lattice spacing $a$ we compare the physical value of the Wilson flow parameter $w_0 = 0.1715(9) \mathrm{fm}$ \cite{w0phys} with its value determined on each ensemble as shown in Table \ref{configs}.

Statistical errors on our results are mostly dominated by the error on $w_0/a$ as used for conversion to physical units. The plots of our results do not include the error on the physical value of $w_0$ since it is correlated between points; it is added later as a systematic error where required.

\section{Results}
\begin{figure}
	\begin{center}
	\includegraphics[width=0.8\columnwidth]{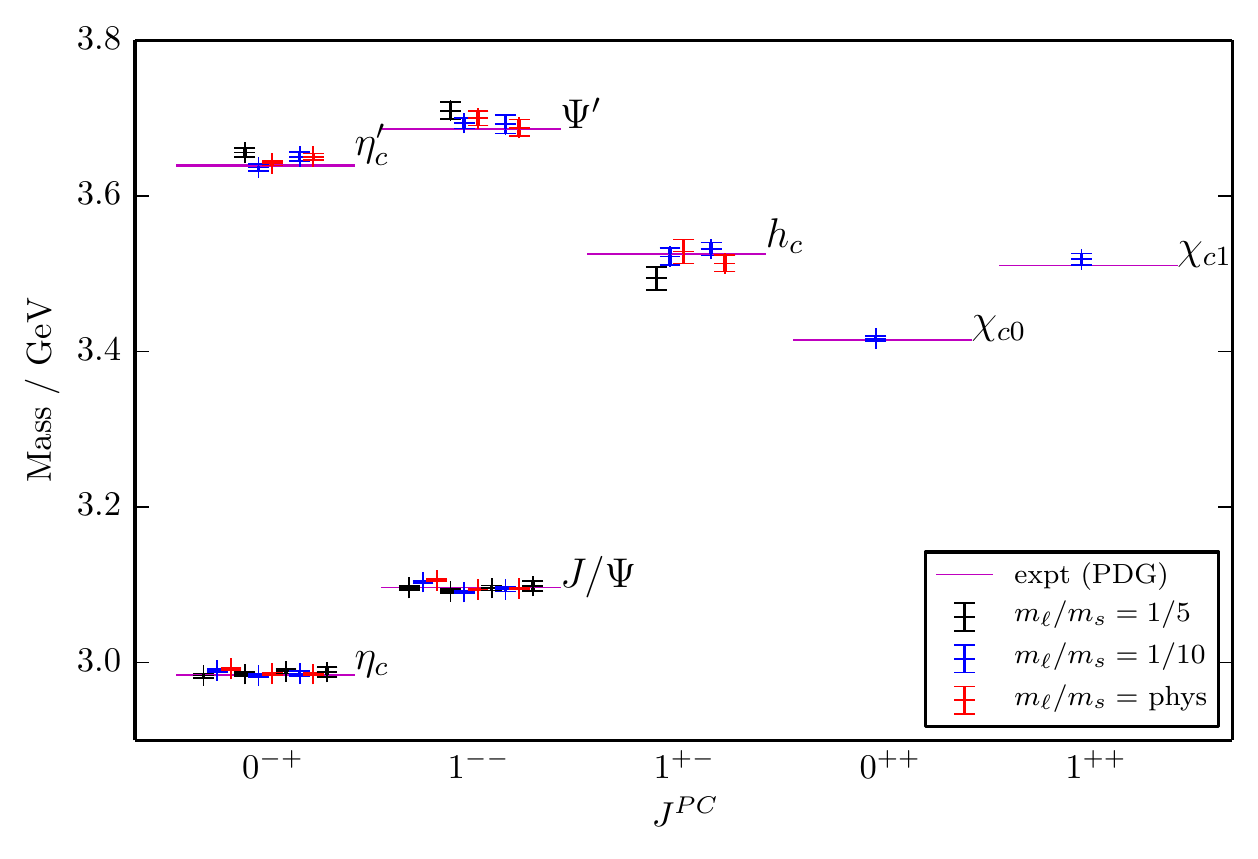}
	\caption{The spectrum of charmonium as computed on each of the ensembles listed in Table \protect\ref{configs}. Individual determinations of each mass are plotted in order of decreasing lattice spacing from left to right. The $\eta_c$ and $J/\Psi$ masses were calculated on all 10 ensembles listed in Table \protect\ref{configs}, and the $\chi_{c0}$ and $\chi_{c1}$ masses have only been computed on a single `coarse' ensemble. The $\eta_c'$, the $\Psi'$ and the $h_c$ masses were computed on the three `coarse' as well as two of the `fine' ensembles.}
	\label{spectrum}
	\end{center}
	\end{figure}

The computed charmonium spectrum can be seen in Figure \ref{spectrum}. Our bare lattice charm masses $am_c$ were tuned by fixing to the mass of the $\eta_c$, and it can be seen from the plot just how well-tuned they actually are.

\subsection{Spin-Averaged 2S--1S Splitting}
\begin{figure}
	\begin{center}
	\includegraphics[width=0.8\columnwidth]{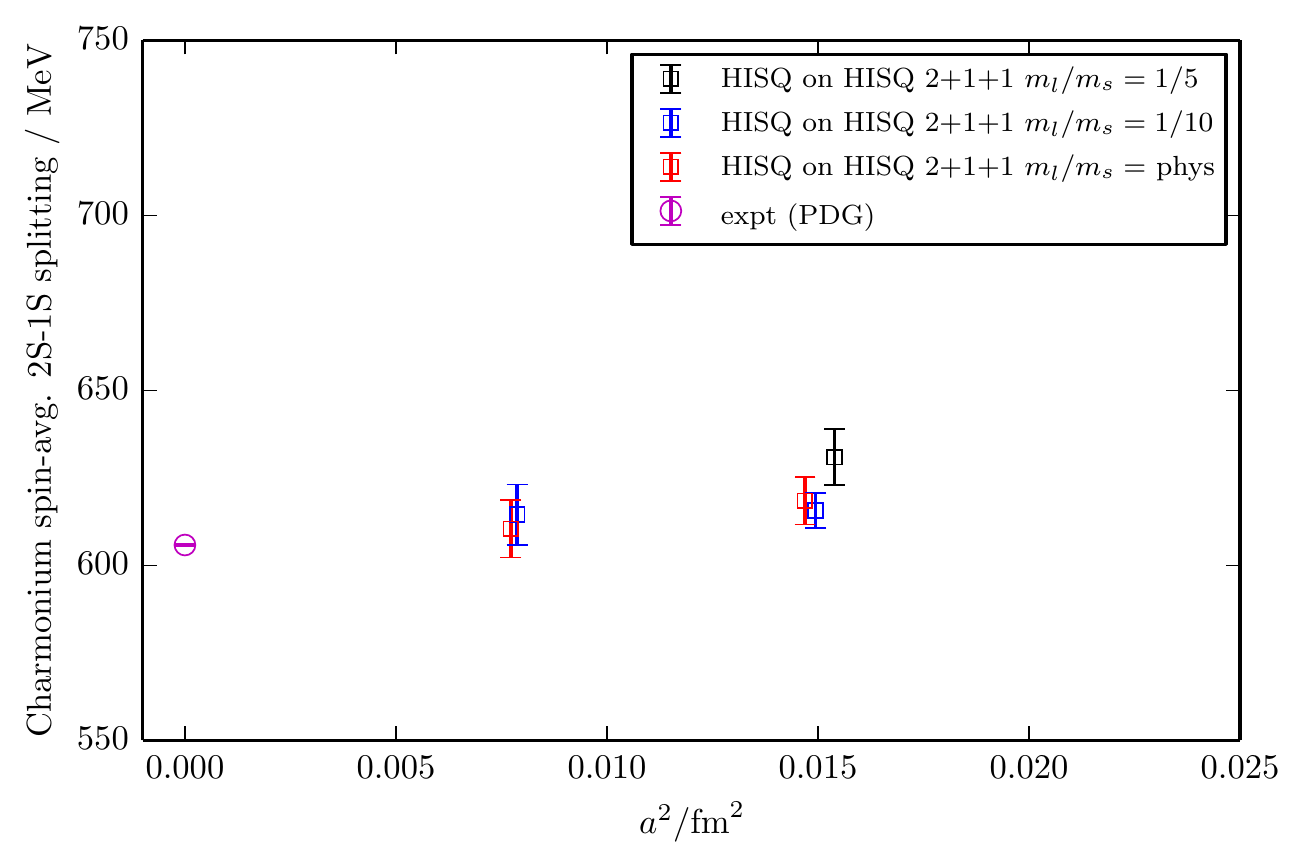}
	\caption{Preliminary results for the spin-averaged 2S--1S splitting of charmonium as determined on `coarse' and `fine' lattices. The magenta point at zero lattice spacing represents the experimental value.}
	\label{2s1s}
	\end{center}
	\end{figure}

A plot of the spin-averaged 2S--1S splitting is shown in Figure \ref{2s1s}. These preliminary results appear to be consistent with the experimental value in the continuum limit, and thus we do not appear to observe the same discrepancy as in the Fermilab study \cite{milc2s1s}.

\subsection{$J/\Psi - \eta_c$ (Hyperfine) Splitting}
\begin{figure}
	\begin{center}
	\includegraphics[width=0.8\columnwidth]{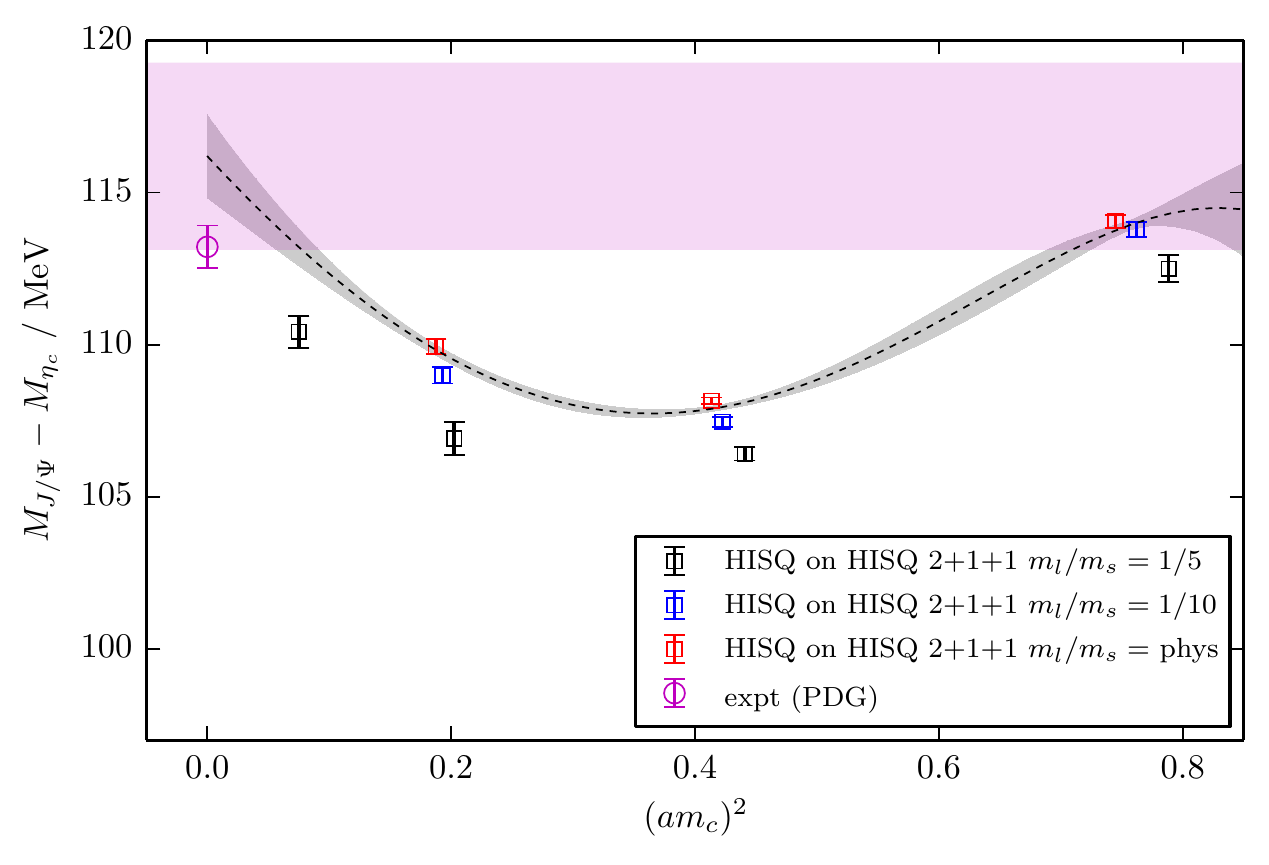}
	\caption{The hyperfine splitting of charmonium as determined on a range of ensembles from `very coarse' to `superfine'. The grey band indicates the fitted curve at the physical light sea quark mass, and the magenta band shows our final result in the continuum limit, including both statistical and systematic errors. Note the range of the y-axis scale.}
	\label{hyperfine}
	\end{center}
	\end{figure}

We have been able to obtain very accurate results for the hyperfine splitting on ensembles with 4 different lattice spacings, and with 3 different light quark masses in the sea, which are plotted in Figure \ref{hyperfine}. Since we have such a wide range of data, we attempt a continuum fit to the form:
	\begin{equation*}
		p\Big(1.0 + A_1 x + A_2 x^2 + A_3 x^3 + A_4 x^4 + A_5 x^5 + \chi_1 \delta_m (1.0 + \chi_{a^2} a^2) + \chi_2 \delta_m^2\Big)
	\end{equation*}
where the $A$ and $\chi$ terms are coefficients to be determined by the fit. $p$ is the physical value of the hyperfine splitting. The terms in $x = (am_c)^2$ model dependence on discretisation errors, and the terms in $\delta_m$ model the sea-quark mass dependence (clearly seen on the plot). This dependence is partly from the use of $w_0/a$ to determine the lattice spacing. $\delta_m$ is the mistuning of the sea quark masses --- this is very small on the physical point lattices.

The prior on the physical value is taken as $p = 110 \pm 20 \text{ MeV}$, and the fit values at the physical point are shown by the grey band. Our continuum result is
		\begin{center}
			$116.2 \pm 1.4 \text{(stat.)} \pm 2.8 \text{(sys.)}$ MeV
		\end{center}
which is shown on the plot as the magenta band. The width of this band includes statistical errors as well as the error on the physical value of $w_0$ as mentioned before; however, it is dominated by the inclusion of a $\pm 2.5 \text{ MeV}$ systematic error to account for $\eta_c$ annihilation effects \cite{hisq}.

This result compares favourably with the current PDG value of $113.2(7)$ MeV \cite{charmonium-pdg}, and also agrees well with our previous calculation on lattices with 2+1 flavours of quarks in the sea \cite{gordon}.

\subsection{$h_c - J/\Psi$ Splitting}
\begin{figure}
	\begin{center}
	
	\begin{subfigure}[b]{0.49\columnwidth}
		\includegraphics[width=\columnwidth]{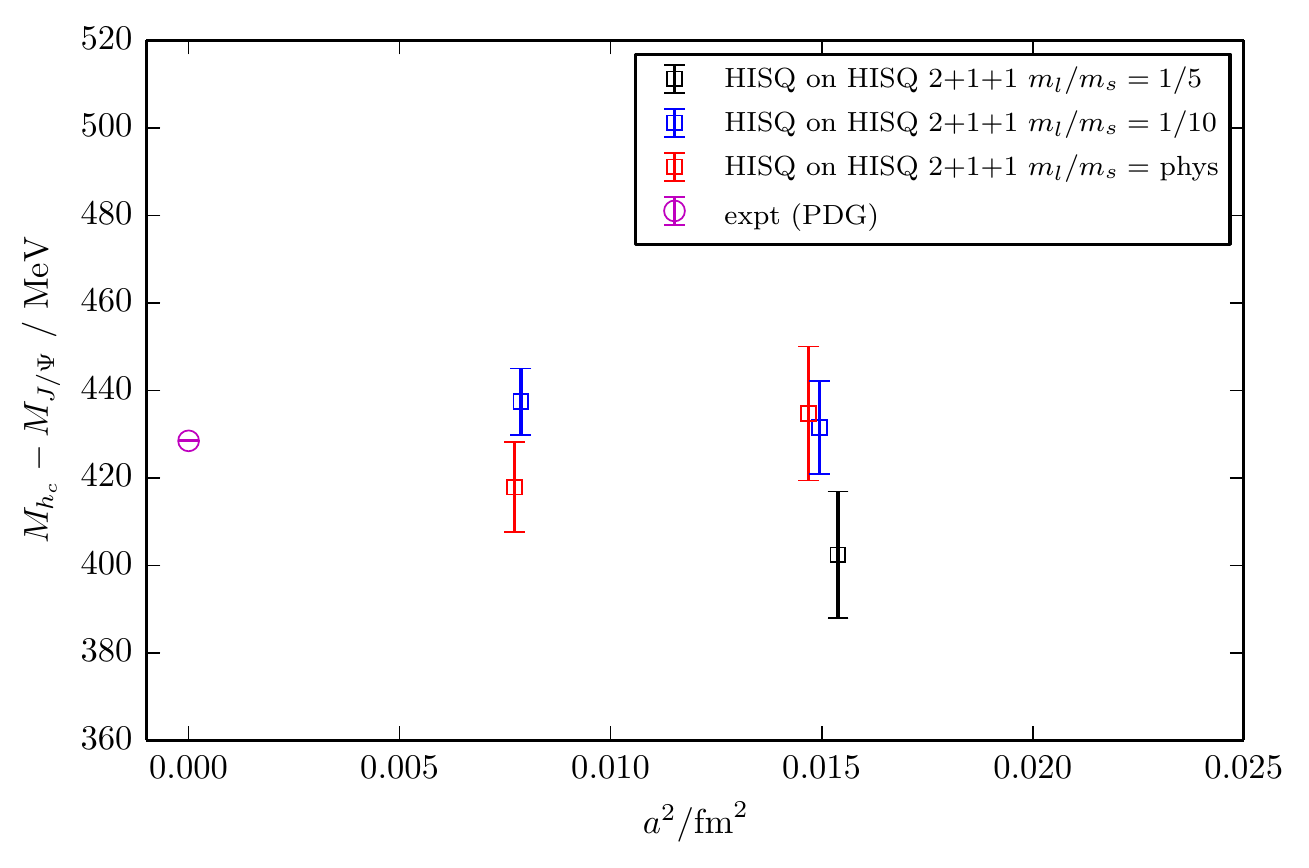}
		\caption{}
		\label{vecsplit}
	\end{subfigure}
	\hfill
	\begin{subfigure}[b]{0.49\columnwidth}
		\includegraphics[width=\columnwidth]{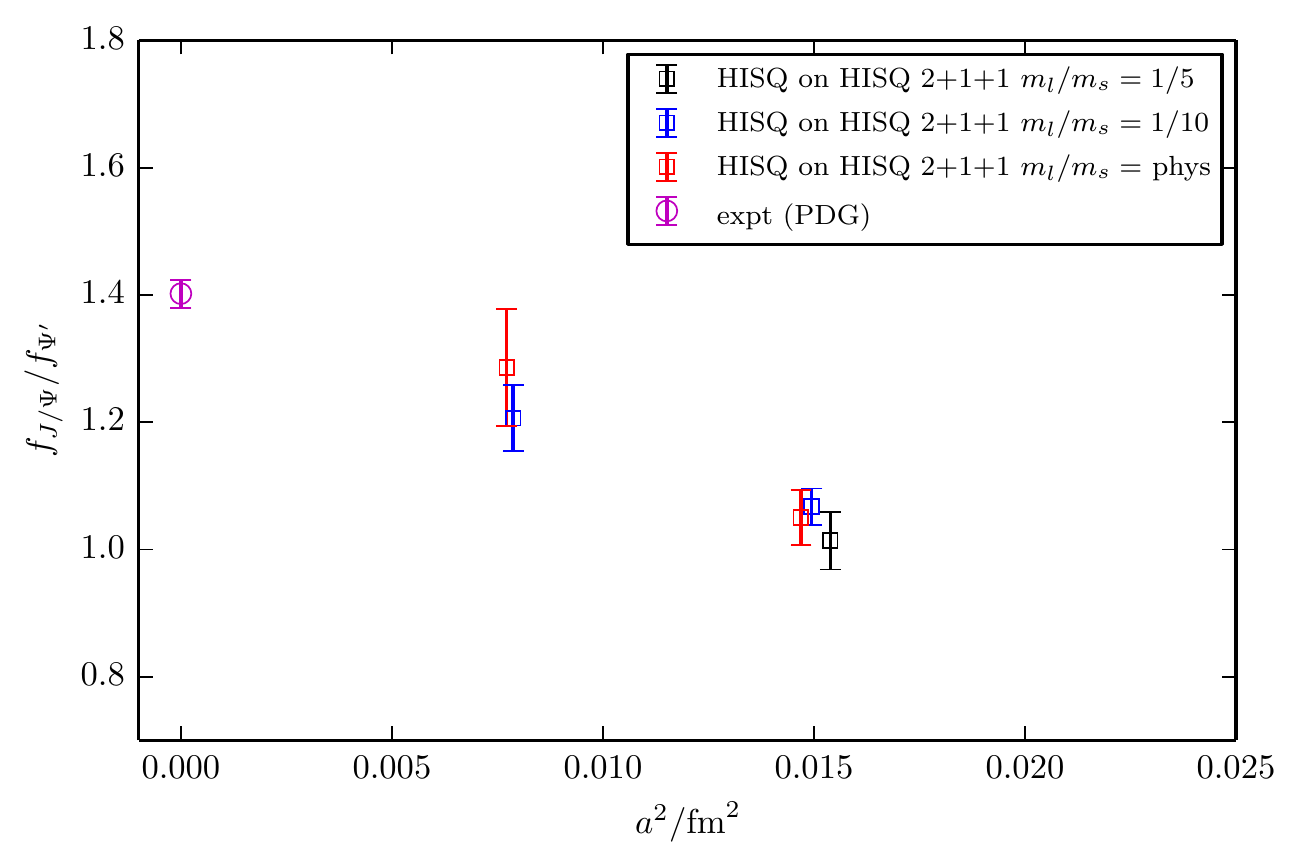}
		\caption{}
		\label{vecratio}
	\end{subfigure}
	
	\caption{Preliminary results for (a) the energy difference between the $h_c$ and the $J/\Psi$, and (b) the ratio of the decay constants of the $J/\Psi$ and the $\Psi'$, as determined on `coarse' and `fine' lattices. The magenta points at zero lattice spacing represent the values derived from experimental measurements.}
	\label{doublefigure}
	\end{center}
	\end{figure}

In Figure \ref{vecsplit}, we plot preliminary results of the energy difference between the $h_c$ and the $J/\Psi$. These also appear consistent with the experimental value in the continuum limit, but more data is needed to make a conclusive statement.

\subsection{Ratio of Vector Decay Constants}
Determining the decay constants of vector states requires knowledge of a renormalisation factor $Z$. However, we can determine ratios of decay constants without requiring this quantity, since the $Z$-factor cancels upon taking the ratio. Preliminary results for the ratio of the decay constants of the $J/\Psi$ and the $\Psi'$ are plotted in Figure \ref{vecratio}, and appear to be consistent with the experimental value in the continuum limit. It is clear that there are significant discretisation effects in this quantity.

\section{Outlook}
We have successfully determined the energies of multiple low-lying states in the charmonium system through the use of smeared source and sink operators with HISQ quarks in lattice QCD. More data is needed to finalise continuum and chiral extrapolations for the spectrum, so an extension of these methods to superfine lattices is underway.

Methods of obtaining the renormalisation constant $Z$ for the vector states are also under development. This will allow us to obtain individual values for the $J/\Psi$ and $\Psi'$ decay constants, rather than the ratio reported above.

A key conclusion is that we find a 2S--1S splitting consistent with experiment.

\acknowledgments
\noindent We extend many thanks to Carleton DeTar for useful discussions, and are grateful to the MILC collaboration for the use of their gauge configurations and code. Our calculations were done on the Darwin supercomputer as part of STFC's DiRAC facility jointly funded by STFC, BIS and the Universities of Cambridge and Glasgow.

\end{document}